\documentclass[11pt]{article}
\usepackage{amsmath, amssymb, amsthm}
\usepackage{graphicx}
\usepackage{epigraph}
\usepackage{times}
\usepackage{comment}
\usepackage[margin=1in]{geometry}
\usepackage{graphicx}
\usepackage{float}  
\usepackage{caption}  
\usepackage{microtype}  
\usepackage{amsmath, amssymb, amsthm}
\usepackage[colorlinks=true, linkcolor=blue, citecolor=blue, urlcolor=blue]{hyperref}
\usepackage{cleveref}  
\usepackage{url}       
\usepackage{booktabs}  
\usepackage{enumitem}  

\title{The First Compute Arms Race: the Early History of Numerical Weather Prediction}
\author{Charles Yang\thanks{Email: contact@charlesyang.io}}
\date{March 2025}

\begin{document}

\maketitle

\begin{abstract}

This paper traces the global race to apply early electronic computers to numerical weather prediction in the decades following World War Two. A brief overview of the early history of numerical weather prediction in the United States, United Kingdom, Sweden, Canada, and Japan is provided. Three critical factors that shaped the development of a national numerical weather prediction are identified: compute capabilities, institution building and state capacity, and talent. Several generalizable lessons are identified with a lens towards modern-day development of national strategies to leverage AI to accelerate scientific competitiveness.

\end{abstract}

\section{Introduction}
\epigraph{Everything is ready, except for the east wind}
{\textit{Zhuge Liang}}

The ability to predict the weather has long been critical to human society: for agriculture, for trade, for warfare. 
In the epic of China's Three Warring Kingdoms, legendary military strategist Cao Cao is defeated during the naval Battle of the Red Cliffs due to an unexpected eastern wind.\footnote{\url{https://apps.dtic.mil/sti/tr/pdf/AD1019690.pdf}}
In more recent western history, the timing of the D-Day invasion during World War Two and the invasion's success is in part due to the superior weather forecasting of the UK Meterological Office ("Met Office") compared to their German counterparts. The German assessment that weather over the channel would be too poor for two more weeks, meant fewer troops were sent to the French front and Rommel returned to Berlin to celebrate his wife's birthday.
\footnote{\url{https://www.metoffice.gov.uk/binaries/content/assets/metofficegovuk/pdf/research/library-and-archive/library/publications/factsheets/factsheet_21-met-office-history-and-timeline_2024.pdf}}
Accurate weather prediction continues to be critical for modern society today, to mitigate natural disasters, to support agricultural production, and to operate a power system increasingly reliant on intermittent renewable energy. 

Yet even during World War Two, weather forecasts still relied primarily on expert forecasters judgment. But the war’s end heralded a seismic shift: the advent of computing machines aand the introduction of Numerical Weather Prediction (NWP), which used computer-solved equations to make weather forecasts. This paper examines the early history of numerical weather prediction in the post war years and the global race to build the largest computers for running NWP, as a historical case study to inform national strategies to advance scientific competitiveness with AI.

\section{The Early Days of Computing}

The idea of Numerical Weather Prediction (NWP) i.e. that future states of the atmosphere are determined by the present state and can be predicted through the use of physics equations, had first been theorized in 1904 by Vilhelm Bjerknes. In 1922, Lewis Fry Richardson outlined the concrete steps to realizing NWP but because all his calculations were carried out by hand and used large time-steps, his predictions were inaccurate and showed unrealistic weather dynamics.\footnote{\url{https://www.cambridge.org/core/books/weather-prediction-by-numerical-process/209AB84257409CF1BB624F97EC9CCA79}}

While long theorized, World War Two supercharged the development of automated computing machines. IBM developed the Harvard Mark I in 1944 to calculate naval gunnery tables. It was subsequently  used by von Neumann for the Manhattan Project. Separately, engineers at the University of Pennsylvania were funded by the US Army to build the ENIAC in early 1943. It was completed in 1946  and moved to Aberdeen Proving Ground Maryland, where it was later used for calculations in designing a hydrogen bomb. Similarly, the United Kingdom had developed the Colossus, an electronic programmable computer used for cracking German encryption (whose fame was perhaps overshadowed by the electromechanical Bombe used for cracking the Enigma). 

After the war, John von Neumann, who had been a major advocate for the role of computing machines during the war, organized the Electronic Computer Project at the Institute for Advancecd Studies (IAS) at Princeton University, with the goal of building a far more powerful electronic computer than the first generation of computers. Von Neumann, with the charismatic Swedish-American Carl Rossby and others, also received funding from the Office of Naval Research to start a Meteorology Group at IAS. Later that year, Rossby organized a meteorology conference at IAS, which brought together meteorologists interested in NWP, including Jules Charney who would end up leading the weather forecasting effort at IAS.\footnote{\url{https://www.jstor.org/stable/26223121}} The Electronic Computer Project developed three application tracks for which this computer would eventually be used: engineering, numerical mathematics, and weather forecasting, the latter of which was led by Jules Charney. 

While the ``IAS machine" was delayed and didn't finish development until 1952, Charney was able to work with the US Weather Bureau to obtain time on the ENIAC machine, now located in Maryland. Working 33 days and nights, the team demonstrated the first Numerical Weather Prediction using the ENIAC in their landmark paper "Numerical Integration of the Barotropic Vorticity Equation" by Charney, Fjørtoft, and von Neumann\footnote{\url{https://maths.ucd.ie/~plynch/eniac/CFvN-1950.pdf}} (referred to hereafter as CFvN).  While CFvN reported taking 24 hours to produce a 24-hour forecast, improvements in computing power could help realize Richardson's dream of advancing compute prediction faster than the weather. Another challenge identified was data input and programming via punch cards, which continued to be a major bottleneck for realistic time-to-result for NWP into the 1950's. \footnote{Indeed, CFvN themselves reported that with their experience, they thought they could achieve a 50\% reduction in time-to-result just from operationalizing data input} \footnote{An analogous bottleneck today is in \href{https://x.com/JasonRute/status/1871188531588026447}{auto-formalization for automated theorem proving}} Critically, the CFvN paper showed surprisingly good forecasting results with a barotropic model, which makes stronger assumptions about weather patterns than a baroclinic model.\footnote{Specifically, a barotropic model assumes fluid density only depends on pressure, whereas a baroclinic model includes temperature and pressure} This would turn out to be circumstantial good luck, but the positive result helped bolster support for operationalizing NWP in the US. 

This result catalyzed the international NWP community, demonstrating for the first time that computers could solve equations to predict the weather.

\section{The Race Afoot}

\subsection{US}
The landmark 1950 CFvN paper, including its surprisingly strong result, was taken seriously by the U.S. Joint Meteorological Committee (JMC), which was under the Joint Chief of Staffs and included the leaders of the US weather agencies: the US Weather Bureau (later renamed National Weather Service), the Air Weather Service, and the Naval Weather Service. They commissioned a subcommittee in late 1952 to investigate the potential of NWP. By summer 1953, the subcommittee returned its recommendation that a joint unit be stood up to operationalize NWP, while tempering expectations that operationalizing NWP would be slow, difficult, and needed to happen concurrently with daily numerical forecasters.

The JMC leadership of all 3 agencies recognized the potential and urgency and actioned the subcommittee's recommendations. On July 1, 1953, the JMC created the Joint Numerical Weather Prediction Unit (JNWPU), a unique unit staffed equally from all three agencies, with civilians working alongside military staff. Early models were developed on the IBM 701 at the IBM headquarters in late 1954, until an IBM 701 was procured specifically for the JNWPU and installed in 1955. Operational NWP was realized and forecasting products were being sent to analysts by that year. It is notable that George Cressman, the director of the JNWPU, would later become head of the National Weather Service in 1965. 

Early on, the JNWPU made the decision that instead of focusing on developing tools for forecasters, they would move to operational NWP production i.e. making 24-hour NWP based forecasts daily to send to forecasters. Despite the fact that the CFvN paper set expectations too high and early NWP results were disappointing, the decision to operationalize helped concentrate resources early on key challenges  e.g. the need for automatic data handling and allowed for later success. \footnote{\url{https://journals.ametsoc.org/view/journals/wefo/4/3/1520-0434_1989_004_0286_honwpa_2_0_co_2.xml}} By 1958, the S1 score (a measure of weather forecasting accuracy) from operationalized NWP demonstrated began to demonstrate the first improvements from its baseline. By 1960, NWP was starting to outperform human forecasts. 

\begin{figure}[h]
\centering
\includegraphics[width=\textwidth]{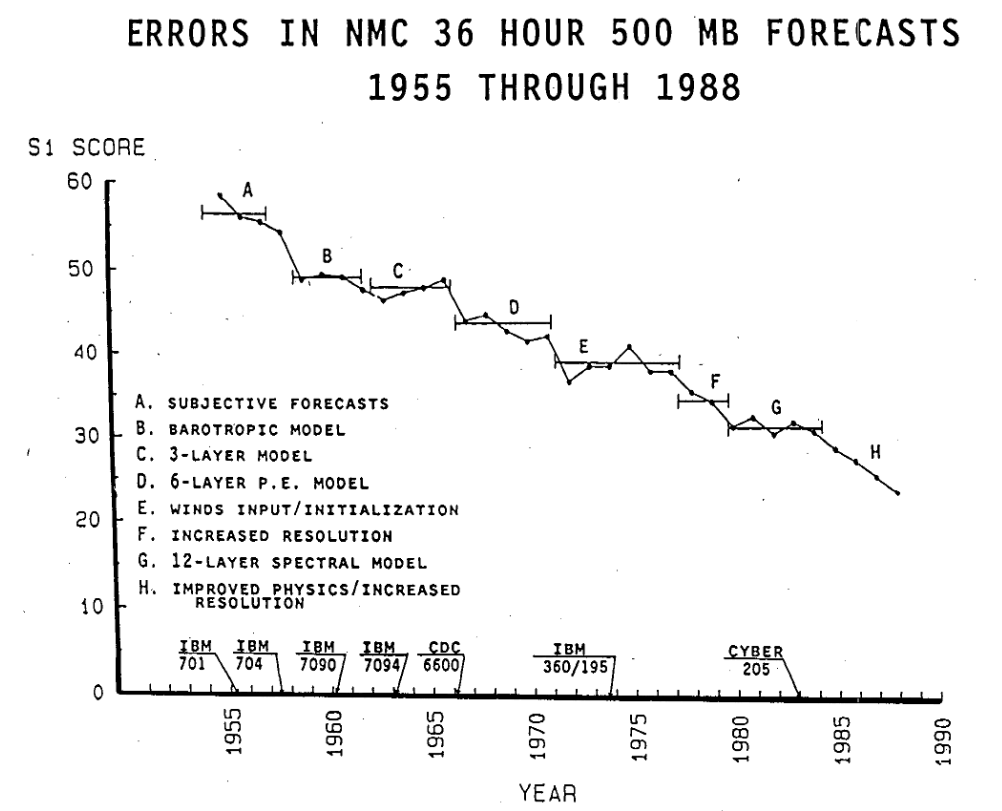} 
\caption{Compute at US NWS. Figure from \href{https://journals.ametsoc.org/view/journals/wefo/4/3/1520-0434_1989_004_0286_honwpa_2_0_co_2.xml}{here}}
\end{figure}

\subsection{UK}
The UK was perhaps the only other country besides the US with the expertise and compute capacity needed to develop NWP, given the Met Office's huge growth during the war and the experience the British gained with computers during the war. However, the UK lacked the same domestic computing industry that the US had in champions like IBM. Indeed, the first computer the UK Met Office (UKMO) borrowed was from Lyons, an enterprising catering shop that had built their own computer to optimize grocery routes. Later, the Met Office used time on Manchester University's Ferranti Mark I, which was developed by Ferranti but modelled after Manchester's Mark I. It was not until 1959 that the UK Met Office was given the funding to acquire their own computer, the Ferranti Meteor. They upgraded to the English Electric KDF9 in 1965. The Met Office was also likely impeded by the UK's long-held desire to support their own domestic champions to compete with IBM.\footnote{For more, see \href{https://www.youtube.com/live/EkTHDgYTh64?}{here}}

In addition to being compute limited, the UK also happened to be unlucky. CFvN used a barotropic model, which made stronger assumptions but also simplified calculations. It so happened that those assumptions were more realistic for weather over the US eastern seaboard than the more complex atlantic weather over the UK.\footnote{Indeed, \href{https://maths.ucd.ie/~plynch/Publications/ENIAC-BAMS-08.pdf}{CFvN} themselves were surprised that a barotropic model could work so well.} Reginald Sutcliffe, who was director of the NWP research program at UKMO,  also favored a baroclinic approach\footnote{fluid density depends on both pressure and temperature}, but this was more computationally complex. As a result, the UK was left with the worst of both worlds: leadership that chose a more compute intensive model while being compute constrained at the same time. Despite making some progress following the 1950 CFvN result on baroclinic approaches, progress stagnated as UKMO continued to need to borrow Manchester's computers on the weekends. From 1956-1959, little progress in NWP was made until UKMO acquired their own computer, despite the US's operationalization of NWP through the JNWPU in 1955.\footnote{It was also during this time that several women meteorologists in the UKMO were noted for their work in this space including being included as co-authors: Mavis Hinds, Vera Huckle, Joe Whitlam, and Margaret Timpson. Mavis and Vera were also inducted into the Royal Meteorological Society}. One retrospective describes the UKMO's double dilemma:

\begin{quotation}

Meteorological centres that a priori ruled out the
barotropic model as a basis for NWP were left with no
other alternative than a baroclinic solution. This would
unavoidably tax the computer’s limited resources,
which made compromises with the computational area
necessary.

\href{https://www.cambridge.org/core/journals/meteorological-applications/article/abs/early-operational-numerical-weather-prediction-outside-the-usa-an-historical-introduction-part-iii-endurance-and-mathematics-british-nwp-19481965/E47615FC785AFB4A79CF731A17CC103B}{Persson, 2005})

\end{quotation}

It is also worth noting that Sutcliffe considered himself first and foremost a forecaster\footnote{"I am happy to claim membership in the forecasters group, once a forecaster always a forecaster." (\href{https://www.cambridge.org/core/journals/meteorological-applications/article/abs/early-operational-numerical-weather-prediction-outside-the-usa-an-historical-introduction-part-iii-endurance-and-mathematics-british-nwp-19481965/E47615FC785AFB4A79CF731A17CC103B}{Sutcliffe, 1956})}. The UKMO, like all meteorological centers at the time, faced the Kuhnian challenge computers posed: how will human forecasters relate to computer forecasts? While some were excited, Sutcliffe was not personally enthused despite heading the NWP program at UKMO:

\begin{quotation}
We are taught to look for the day when machines will calculate the future weather with a monotonous degree of success, but if that day comes one satisfying profession will be lost to man and we must look elsewhere. There will be little more joy in the trade than there is in the repetition of the multiplication table.

\href{https://www.cambridge.org/core/journals/meteorological-applications/article/abs/early-operational-numerical-weather-prediction-outside-the-usa-an-historical-introduction-part-iii-endurance-and-mathematics-british-nwp-19481965/E47615FC785AFB4A79CF731A17CC103B}{Sutcliffe 1956}

\end{quotation}

It was not until a new UKMO director, John Mason, was appointed in 1965 did the UKMO start to change. Mason felt that the UKMO "had ‘an enormous potential’ but was rather bureaucratic and its
staff on the operational side was lacking in confidence" \footnote{\url{https://www.cambridge.org/core/journals/meteorological-applications/article/abs/early-operational-numerical-weather-prediction-outside-the-usa-an-historical-introduction-part-iii-endurance-and-mathematics-british-nwp-19481965/E47615FC785AFB4A79CF731A17CC103B}}. Despite staff protests that they needed 6-12 more months to safely operationalize NWP, Mason pushed forward with a press conference to announce operationalized NWP forecasts 1 month from when he started as director. The staff were able to pull off operationalization successfully in time in and the press conference on November 2 1965 was a huge success. There was again an element of luck - the accurate forecast produced by NWP that day was the best forecast for the next several months.

\subsection{Sweden}

Sweden was surprisingly an early lead in advancing NNWP, being the first country to operationalize real-time, +72 hour NWP during military exercises in 1954\footnote{\url{https://rmets.onlinelibrary.wiley.com/doi/pdf/10.1017/S1350482705001593}} and briefly owned the world's fastest supercomputer in 1953\footnote{\url{https://historyofinformation.com/detail.php?id=2073}}.

Sweden's early leadership role in advancing NWP is due in large part to Carl Gustaf Rossby, a Swedish-American and eminent meteorologist. He was the first head of the Department of Meteorology at the Massachusetts Institute of Technology \footnote{\url{https://www.nature.com/articles/1801166a0.pdf}}. During World War Two, he served as an assistant chief of research at the US Weather Bureau, helping train the first weather service for civil aviation. He also served as president of the American Meteorological Society from 1944-1945 and was chair of meteorology at the University of Chicago. \footnote{\url{https://history.aip.org/phn/11608040.html}} \footnote{\url{https://www.nature.com/articles/1801166a0.pdf}} He was most known for his characterization of the eponymous Rossby waves\footnote{\url{https://en.wikipedia.org/wiki/Rossby_wave}}, which are planetary undulating waves driven by the polar jet stream. 

After the war, driven in part by temperament and origin and by the encouragement of the Swedish government which created a chair position at Stockholm University, Rossby returned to Sweden. He immediately began hiring and inviting his former US and other international colleagues to come to Stockholm as visiting fellows, which many of them did. He also launched a new journal, the Tellus, to serve as a counterweight to the 2 other European theoretical meteorology journals. (Rossby had founded the Journal for Meteorology in the US previously.)\footnote{\url{https://maths.ucd.ie/~plynch/eniac/CFvN-1950.pdf}} It was in the Tellus that the groundbreaking result from Carney, von Neumann, was published, demonstrating that the ENIAC could solve numerical integration problems for weather forecasting. \footnote{\url{https://maths.ucd.ie/~plynch/eniac/CFvN-1950.pdf}}

At the same time, the Swedish government had become quite interested in the computer development in the US. The government created the Swedish Board of Computer Machinery (SBCM), a government state-owned enterprise, to procure machines from US companies. But when the State Department signalled that these computer machines would be tightly export controlled in 1948 (as a general policy and not specific to Sweden), SBCM pivoted to building their own computers. The BESK was SBCM's second computer and heavily influenced by von Neumann's Princeton machine. It was operational by December 1953 - discussion around its installation was dominated by its potential utility to meteorology. 

To truly operationalize NWP, Rossby needed a sponsor. Time on the BESK was expensive and to run daily operations would require more funding. Rossby originally approached the Swedish Meteorological and Hydrological Institute (SMHI), Sweden's meteorological office, about implementing an operational NWP. But the SMHI leadership felt some degree of skepticism towards NWP, believing something this experimental should be left to universities. Rossby then approached the Military Weather Central (MVC) under the Swedish Air Force about sponsoring operational NWP. The leader of the MVC, Oskar Herlin, was a practical man who had felt for some time that progress in conventional techniques for +1 to +24hr forecasts had stagnated. It was in this pessimistic environment that the charismatic Rossby approached Herlin. Given the relatively small cost compared to the overall military budget, Herlin was easily convinced of its importance and was able to convince the Swedish Air Force as well.

In September 1954, an opportunity presented itself: 45,000 troops in central Sweden were conducting military maneuvers, including testing nuclear protection equipment. Real-time forecasts of upper air movement, which is strongly determinative of the trajectory of the simulated nuclear fallout, would therefore be critical. The operationalized NWP forecasts turned out to be tremendously successful compared to subjective forecasts. The success and buzz generated by BESK further influenced the appointment of new leadership at SMHI, who appropriately updated their views on the potential of computing for meteorology. While SMHI took a greater involvement in NWP going forward, it did not operationalize NWP until 1961 on the BESK and would not be fully responsible for all NWP production in Sweden until 1973. 

Because Sweden was one of the first to seriously operationalize NWP, they were also one of the first to deal with the human challenges of integrating NWP into the practice of weather forecasting. While forecasters could openly discuss any discrepancies between their forecasts, how should human forecasters address discrepancies with NWP forecasts? To address this, the Swedes developed a "User Guide" for human forecasters. The anonymous guide recommended generally accepting NWP forecasts, given its known accuracy over human forecasting at that point, but identified several areas where NWP forecasts were known to be problematic. In addition, NWP forecasts could change drastically from one day to the next. This created a messaging problem for the forecasting community who had to explain their forecasts to the public - at least one of the forecasts had to be wrong. The Guide recommended providing continuity where possible and explaining any discrepancies.     

By the 1970's, discussions around the European Centre for Medium-Range  Weather Forecasts (ECMWF) had already started and Sweden diverted resources towards that effort. It is notable that three of the senior figures starting ECMWF began their careers working with Rossby's NWP project 25 years ago. It also speaks to Rossby's international character that only one of them was Swedish. \footnote{\url{https://rmets.onlinelibrary.wiley.com/doi/pdf/10.1017/S1350482705001593}}

And what happened to Sweden's nascent compute market? Here as in other cases, the liberalization of U.S. export controls in the 1960's and the introduction of competition from IBM played a key role in crushing domestic competition.\footnote{The full history of Sweden's attempts to build their own domestic electronic computer can be found \href{https://ieeexplore.ieee.org/abstract/document/1549794/similar}{here (article)}, \href{https://www.youtube.com/watch?v=QJPMIjVGzd0}{here (video essay)}, and \href{https://www.youtube.com/watch?v=UXSBGjWSG7Y}{here (second video essay)}}.  

\subsection{Japan}

While Sweden's NWP and compute programs advanced quickly because they avoided the worst devastation of World War Two and benefited from Rossby's repatriation, Japan's story is rather the opposite. 

Japan suffered severe economic devastation from World War Two and in 1950, was still under allied occupied governance. But the origin for Japan's NWP can be attributed to Shigekata Syono, who was appointed as a professor meteorology at Tokyo University in 1945. He saw the transformative shift in the field of meteorology heralded by the 1950 Carney paper and helped stand up a NWP group at Tokyo University, with his graduate students and Japan Meteorological Agency (JMA) researchers. They quickly published a series of papers in 1950 on Numerical Weather Prediction, comparing and expanding on the work done by Carney et al.  In particular, one of Soyono's students, Kanzaburo Gambo, began a correspondence with Charney and was invited to spend 2 years at Institute for Advanced Studies (IAS) from 1952-1954. Their quick pace of work came despite limited research funding available in Japan at the time. Many of the graduate students took part-time work as tutors to compensate for lack of funding.

When Gambo returned from IAS, he and Syono launched a monthly seminar series on NWP with Tokyo University and JMA researchers. The NWP group continued an impressive pace of research, including the first application of barotropic numerical models to typhoon tracking. 

Their progress was in spite of limited funding and limited compute. Their greatest sponsor was not the federal government, but a grant in 1954 from one of the country's largest newspapers, the Asahi Press. Similarly, Tokyo University researchers used a Fujitsu electromechanical relay switching computer, which was much slower than the vacuum tube computers that were being developed. It was not until 1959 that JMA received funding for a computer and was able to procure and install a computer (IBM 704). They were able to operationalize NWP that year. JMA did not procure a new computer until 8 years later in 1967 (a HITAC 5020). Limited fiscal spending and headcount growth at JMA also limited the career prospects for talented Japanese graduate students in the NWP group Gambo and Syono created, despite their seminal early work. Combined with a generous US immigration policy for foreign scientists, almost a dozen graduate students from the Syono group ended up emigrating to US universities during the 1950s and early 1960s. \footnote{\url{https://journals.ametsoc.org/view/journals/bams/74/7/1520-0477_1993_074_1351_mftuot_2_0_co_2.xml}}

Syono's efforts culminated with the 1960 international NWP symposium hosted in Japan, which was "epoch making" in recognizing the maturity of the field and the efforts in Japan around NWP, despite the difficult post-war environment. 

\begin{figure}[h]
\centering
\includegraphics[width=\textwidth]{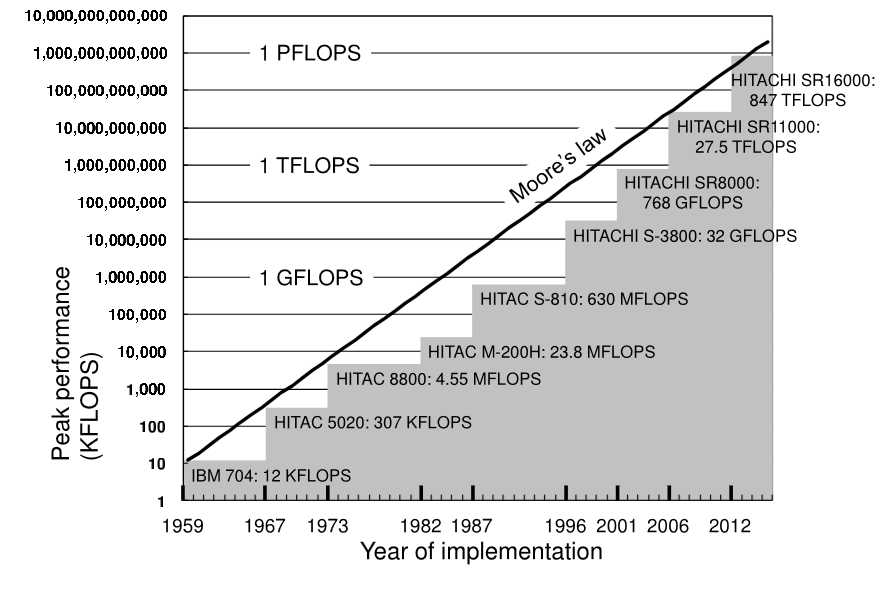} 
\caption{Compute at JMA. Figure from \href{https://www.jma.go.jp/jma/jma-eng/jma-center/nwp/outline2013-nwp/pdf/outline2013_01.pdf}{here}}
\end{figure}

\subsection{Canada}
Unlike the US, Canada lacked ready access to computers and even worse, the leadership of the Canadian Meteorological Centre (CMC) "really did not believe in NWP".\footnote{\url{https://rmets.onlinelibrary.wiley.com/doi/pdf/10.1017/S1350482705001751}} It was only because of the persistent belief of a research director, Michael Kwizak, who pushed for the resources and talent needed, that Canada was even able to achieve operational NWP. The CMC had to purchase time on Canadair's computer until they acquired their own computer in 1963, a Bendix G-20\footnote{Also an American company based in Los Angeles.}, and achieved operational NWP later that year. Later in 1967, they acquired an IBM 360/65. 

\section{Common Trends and Analogies to AI for Science}

\subsection{Compute}

While it took significant effort to operationalize NWP models on early computers—especially given rapidly evolving data input systems—it quickly became clear that more powerful machines enabled higher model resolution and better dynamical fidelity. In the beginning, electronic computers were built not by corporations but by universities, many of which were staffed by researchers returning from the war eager to push this new technology forward. As machines grew larger and more capable, the global race to acquire them for national weather services required weather bureaus to scale up funding. It is notable that in the U.S., the weather service was one of the first customers for IBM, demonstrating the role the public sector and public capital can play in supporting private sector innovation.\footnote{\url{https://en.wikipedia.org/wiki/IBM_701\#IBM_701_customers}}.

But the push to secure cutting-edge computers clashed with another common policy objective: the desire to cultivate domestic computing industries. For instance, JMA used their first funding to purchase an IBM 704, a competitive computer which brought them closer to parity with the US weather bureau, which had access to compute for much longer. However, in subsequent years, JMA exclusively used Hitachi systems, as part of a broader industrial policy from Japan's Ministry for International Trade and Innovation to build a domestic computing industry capable of competing with IBM.\footnote{\url{https://www.princeton.edu/~ota/disk2/1990/9007/900713.PDF}} Similarly, the United Kingdom \footnote{\url{https://www.youtube.com/watch?v=EkTHDgYTh64}} and Sweden \footnote{\url{https://www.youtube.com/watch?v=QJPMIjVGzd0}} both invested significant amounts of funding and state capacity to building their own compute industries, both of which ultimately failed to successfully compete with American giants like IBM. 

\begin{table}[h]
    \centering
    \begin{tabular}{|c|c|c|c|}
    \toprule
        \textbf{Country} & \textbf{Year Computer Acquired} & \textbf{Computer Model} & \textbf{kFLOPS}\\ 
        \midrule
         Sweden&  1954& BESK & ? \\ 
         US&  1955& IBM 701& ? \\ 
         Japan& 1959 & IBM 704& 12\\ 
        UK& 1959 & Ferranti Meteor & 0.3 \\ 
         Canada & 1963 & Bendix G-20& ?\\ 

    \bottomrule
    \end{tabular}
    \caption{First Dedicated NWP Computer by Country}
    \label{tab:table_one}
\end{table}

Today, the story is remarkably similar. The US continues to maintain an enormous lead in compute with Nvidia GPU's being the default option and the only real contenders are also American companies (AMD GPU, Google TPU, Amazon Trainium). However, China's tremendous rise in development since the 1950's and the aggressive use of US chip export controls means they are also developing their own compute ecosystem e.g. Huawei's Ascend GPU\footnote{\url{https://cset.georgetown.edu/publication/pushing-the-limits-huaweis-ai-chip-tests-u-s-export-controls/}}. Only time will tell how successful China's efforts will be. The UK government also tried to create a national champion out of Graphcore, an AI chip startup, but that effort ultimately failed.\footnote{\url{https://www.wired.com/story/graphcore-uk-ai-champion-scrambling-to-stay-afloat/}} 

The other race is to actually build large-scale compute clusters to develop advanced models. In the case of NWP, we see the importance of government agencies having access to large-scale compute systems, which correlated strongly with their ability to operationalize computational breakthroughs. In this regard, the U.S. has fallen behind today in meteorology: the \href{https://www.ecmwf.int/en/about/media-centre/key-facts-and-figures}{Europeans have a larger supercomputer} (30PF) than \href{https://www.weather.gov/about/supercomputers}{NOAA} (8PF) does today. As a result, U.S. weather models lag in performance to their European and British counterparts.\footnote{\url{https://journals.ametsoc.org/view/journals/bams/104/3/BAMS-D-22-0172.1.xml?tab_body=pdf}} 

Access to AI compute will play an equally important role in accelerating AI for Science today. While the U.S. owns the worlds largest supercomputers today, hosted at the Department of Energy national labs\footnote{\url{https://top500.org/}}, private U.S. companies increasingly own the largest GPU clusters needed specifically for large AI model training. U.S. efforts like the National AI Research Resource (NAIRR)\footnote{\url{https://nairrpilot.org/}} and the Department of Energy's Frontiers in AI for Science, Security, and Technology (FASST) \footnote{\url{https://www.energy.gov/fasst}} are examples of recent federal efforts to increase researcher access to AI compute. But further efforts will be needed to realize the diffusion and effective deployment of AI in scientific domains. Multiple studies have demonstrated the critical role of modern compute capacity in advancing scientific progress. \footnote{See \href{https://papers.ssrn.com/sol3/papers.cfm?abstract_id=4868123}{this study} on the role compute capacity plays in scientific innovation} \footnote{\url{https://arxiv.org/abs/2206.14007}}

\begin{figure}[h]
\centering
\includegraphics[width=\textwidth]{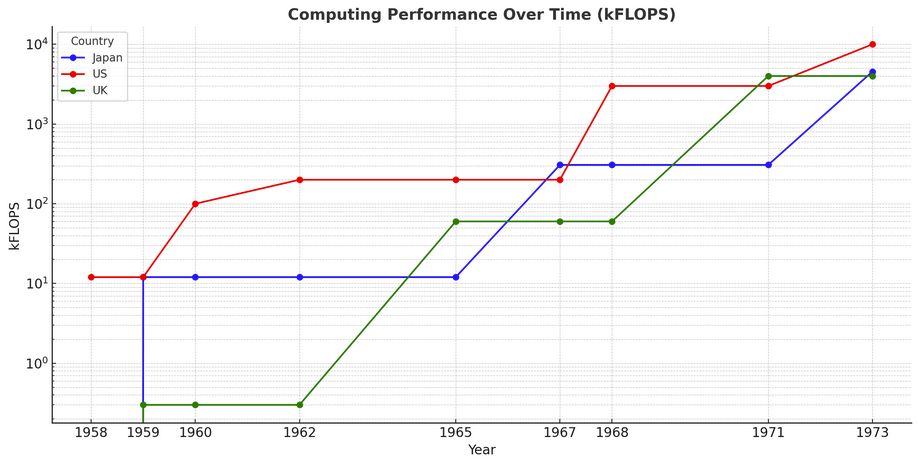} 
\caption{Weather Forecasting Compute Performance of US, UK, Japan}
\label{fig:compute}
\end{figure}

\subsection{State Capacity, Institutions, and Bureaucracy}

\begin{table}[h]
    \centering
    \begin{tabular}{|c|c|c|}
    \toprule
        \textbf{Country} & \textbf{Year Operational NWP Achieved} \\
        \midrule
         Sweden&  1954\\  
         US&  1955 \\ 
         Japan& 1959  \\ 
         Canada & 1963 \\ 
         UK&1965  \\  
         \bottomrule
    \end{tabular}
    \caption{Year Operational NWP Achieved by Country}
    \label{tab:table_two}
\end{table}

Operationalizing NWP required not just the technical workforce and compute, but also significant government investment and buy-in, given weather forecasting's traditional public sector remit. The U.S.'s early leadership in this technology is due in part to the U.S. political and military leadership recognizing the importance of this technology. Despite the difficulties of coordinating three different military and civilian organizations, the U.S. was able to create an efficient structure through the JNWPU, committed to technological innovation despite its disruption to existing institutions, and invested public capital in procuring compute capabilities early. While the U.S. had a domestic compute industry, it required capital for the government to translate that industrial advantage into an actual national capability. Canada and the U.K. did not have that same far-sighted leadership that was willing to invest capital to drive innovation, despite the uncertain impact on human forecasters. Note the strong correlation between when countries acquired their first computer (\autoref{tab:table_one}) and when operational NWP was first achieved (\autoref{tab:table_two}).

Similarly, integrating AI into scientific discovery will require cultural and educational shifts within scientific institutions. The Department of Energy's "AI Jam Session", which hosted 1000 scientists across 9 national labs, is a good example of how institutions can encourage adoption of new technologies.\footnote{\url{https://www.anl.gov/cels/1000-scientist-ai-jam-session}}. But advancing being competitive in AI for Science will be shaped by how quickly science funding agencies and institutions prioritize early investment, incorporate AI for Science into research efforts, and elevate early AI leaders within scientific fields. Institutions must overcome sluggish bureaucracy, kuhnian resistance to new paradigms, and oftentimes source new pools of capital investment - but 

\subsection{Labor Flows and Immigration}
During the post war period, the U.S. occupied an enviable position of relative prosperity compared to the war torn or less developed regions of the word. The stories of Syono's graduate students immigration to the US and conversely Rossby's move to Sweden illustrate the role talent played in determining the geographical development of NWP and computers for weather forecasting. In the particularly early days of building computers and "synoptic" forecasting, the pace of progress in developing new models was rapid and mediated by a small group of talented researchers in universities and agencies across the world.

Today, a similar race for talent is underway in AI. The United States still retains a marginal lead as a destination for top researchers, as of 2024.\footnote{\url{https://archivemacropolo.org/interactive/digital-projects/the-global-ai-talent-tracker/}}  But that lead is increasingly fragile. Anti-immigration sentiment, political instability, and more attractive opportunities abroad are reshaping global talent flows. For example, China’s DeepSeek has emerged as a frontier model laboratory staffed entirely by domestic researchers - many of whom have no academic time abroad. In Canada, Alan Aspuru-Guzik left Harvard in 2018, citing concerns over the U.S. political climate. In his interview with the Harvard Crimson, he describes his reason for moving as "primarily due to his concern over the nation’s political climate, especially fallout from the tumultuous 2016 presidential election".\footnote{\url{https://www.thecrimson.com/article/2018/3/30/chem-prof-to-canada/}} Today, Alan leads the Acceleration Consortium at University of Toronto, which is recognized as a world leader in \href{https://ml4sci.substack.com/p/self-driving-labs}{self-driving lab development} and recipient of Canada's largest single scientific grant\footnote{\url{https://acceleration.utoronto.ca/news/u-of-t-receives-200-million-grant-to-support-acceleration-consortiums-self-driving-labs-research}}. In some ways, Aspuru-Guzik mirrors Rossby's story - a single figure whose relocation sparked institutional catalyzed institutional investment in a new national context (though notably they left the U.S. for very different reasons). In contrast, the U.S. has only very recently started to invest in autonomous experimentation at a much lower funding level\footnote{\url{https://ml4sci.substack.com/p/us-progress-on-self-driving-labs}}.

But it’s not just about national talent pools—it’s about where that talent goes. Today’s leading AI researchers are often in private industry rather than public research agencies. Indeed in meteorology today, it is not a public institution but Google Deepmind that is likely the world leader in developing AI models for weather forecasting.\footnote{\url{https://deepmind.google/discover/blog/gencast-predicts-weather-and-the-risks-of-extreme-conditions-with-sota-accuracy/}} Scientific institutions need to effectively nurture and compete for AI talent, if they wish to remain 

\section{Conclusion}

The early history of computing in weather forecasting offers a revealing case study in how nations convert technological capacity into scientific capability. In the years following World War Two, the United States combined an emerging commercial compute industry with forward-leaning public institutions and a close monopoly of global scientific talent. That alignment enabled the U.S. to lead the world in operationalizing numerical weather prediction and, for a time, to define the frontier of computational science.

Today, the landscape is more complex. The U.S. is no longer a world leader in NWP. While the U.S. still has a world-leading AI compute industry, China is making meaningful and substantial efforts to compete on AI chip development. U.S. political turmoil and the decimation of state capacity may make it difficult for public research institutions to invest in AI for scientific research and are leading to talent outflows from the U.S. If the twentieth century's breakthroughs in NWP were driven by governments willing to bet early on compute, talent, and institutional experimentation, the future of AI for science will demand a similar posture. 

\section{Acknowledgments}
This paper was inspired by the Asianometry's video "\href{https://www.youtube.com/watch?v=OY3AeB-VVO0}{The Computer Revolutionized Weather Forecasting}". Thanks to the organizers of \href{https://www.rabbitholeathon.com/}{rabbitholeathon 5.0} for creating a cozy environment for me to properly settle into this fascinating rabbit hole. 

Thanks also to \href{https://sv.wikipedia.org/wiki/Anders_Persson_(meteorolog)}{Anders Persson}, whose historical research, interviews, and publications documented the stories of the people behind the origin of NWP.

\section{Appendix}

Data for \autoref{fig:compute}

\begin{table}[h]
\centering
\begin{tabular}{|c|c|p{8cm}|}
\toprule
\textbf{Year} & \textbf{kFLOPS} & \textbf{Notes} \\
\midrule
1959 & 12 & IBM 704 \\
1967 & 307 & HITAC 5020 \\
1973 & 4550 & HITAC 8800 \\
1982 & 23800 &  HITAC M-200H \\
\bottomrule
\end{tabular}
\caption{Japan – Computing Performance for Weather Prediction}
\label{tab:japan_kflops}
\end{table}

\begin{table}[h]
\centering
\begin{tabular}{|c|c|p{8cm}|}
\toprule
\textbf{Year} & \textbf{kFLOPS} & \textbf{Notes} \\
\midrule
1955 &  & IBM 701 \\
1958 & 12 & IBM 704 \\
1960 & 100 & IBM 7090 \\
1962 & 200 & IBM 7094 (IBM 7090 was 100, 7094 reportedly up to 2x as fast) \\
1968 & 3000 & CDC 6600 \\
1973 & 10000 & IBM 360/195 \\
\bottomrule
\end{tabular}
\caption{United States – Computing Performance for Weather Prediction}
\label{tab:us_kflops}
\end{table}

\begin{table}[h]
\centering
\begin{tabular}{|c|c|p{8cm}|}
\toprule
\textbf{Year} & \textbf{kFLOPS} & \textbf{Notes} \\
\midrule
1959 & 0.3 & Ferranti Meteor \\
1963 & ? & LEO KDF9 \\
1965 & 60 & Comet \\
1971 & 4000 & IBM 360 \\
\bottomrule
\end{tabular}
\caption{United Kingdom – Computing Performance for Weather Prediction}
\label{tab:uk_kflops}
\end{table}

\end{document}